# Knowledge Enhancement and Mobile Technology: Improving Effectiveness and Efficiency

Siddhartha Paul Tiwari

Google Asia Pacific, Maple Tree Business City, Singapore

Email: sidpaultiwari@gmail.com

*Abstract*

In education, mobile technology creates value on three fundamental pillars: productivity, coordination, and transformation. Mobile apps are becoming an increasingly important aspect of teaching and learning in many countries. The use of mobile applications in education is not only beneficial, but also provides students with an enjoyable and interactive experience. For a mobile product launch to be as successful as it can be, it is imperative that a systematic, precise, controlled, and well-established process is in place, which is controlled, efficient, and well-established. Many educational organizations find themselves in situations where they have to get all departments working effectively and together in order to meet a specific deadline, including marketing, production, and operations, after the organization's product clearance board approves the new product. In many ways, the situation is similar to the software crisis that took place in the middle of the 1970's. As a result of globalization and communication, oftentimes the effects of globalization are amplified because of the vast amount of information that must be shared among project team members. Every educational organization has a unique style and way of doing things, and the project management team is no exception. As a consequence, it is commonplace to see that every education entity has its own particular style and way of doing things. In regards to the creation of a mobile application that is going to function efficiently, it is important to remember that it is extremely important to stick to the strategies and requirements that will yield the best results for the education sector.

## Introduction

Although much progress has been made in the technology sector in the past decade, most educational organizations are still heavily focused on managing legacy mobile applications and systems and reducing costs. Therefore, decision makers are unable to pursue the more strategic objective of creating greater value through the use of analytics, systems, and applications on mobile devices. By focusing on the efficiency and alignment of operations, decision makers can begin to drive innovation, which is the key to unlocking the full potential of mobile technologies in the education sector. Furthermore, Mehdipour, Y., & Zerehkafi, H. (2013) identified the benefits and challenges of mobile learning for the education sector. The goal of this viewpoint is to consider the value creation journey, emphasizing particularly the factors that are needed to unlock the potential of mobile applications and systems.

In the last few years, Covid has been the cause of a worsened economic environment. Educational organizations have been forced to launch cautious products as a result of rising costs and unpredictable markets. It has therefore become increasingly difficult for instructors and overall education sector to meet their expectations. During the process of launching new products, educational organizations can track progress by using time-tracking software and scheduling tools, ensuring the deadlines are not missed. In spite of the project monitoring tools, the launch dates of the mobile products have not been met despite the use of these tools. In light of the unprecedented challenges faced by mobile products launched as a result of Covid, certain points have emerged, including the urgency at the forefront of launching a new product. Mobile product development has to be undertaken at a rapid pace in order to meet the demands of a dynamic competitive environment. There has been demand on product teams to deliver mobile technologies and apps that could be developed, tested, and shipped within a short period of time. Test methods that have been used traditionally are known to be slow and might not keep pace with the fast-paced new world order we are entering.

**How Mobile Technology Creates Value for Education Sector**

The three fundamental pillars on which mobile technology in education creates value are productivity, coordination, and transformation. Additionally, Vázquez-Cano, E. (2014) provides detailed information regarding mobile distance learning in higher education using smartphones and apps. In many countries, mobile apps are becoming an increasingly important part of education. The use of mobile apps in education is not only beneficial, but also provides students with an enjoyable and interactive experience. Various educational mobile apps are designed specifically to target the psychology of students in order to assist students in understanding and assimilating the information from a different perspective. The app facilitates students' understanding of concepts by presenting them with challenging tasks, puzzles, and educational games. Khan Academy and Duolingo are just some of the most popular educational mobile apps available today. It is possible for people without access to education to learn some skills by utilizing mobile applications for educational institutions, even if they don't have access to education. Learning will no longer be restricted to the classroom with the use of educational apps, as students will have the power to take control of their own learning at any time and can challenge themselves. By mastering the concepts they are learning in the classroom through learning apps, students have the chance to benefit greatly from their studies. It is important to note that the use of educational applications is not only restricted to creative learning; they can also be used for accommodating students according to their convenience, since it can be quite exhausting for both students and teachers to deliver a lengthy lecture. These days, educational apps for mobile phones and e-books are the first choice for students who want to learn on their own schedule because they offer students the possibility of learning on the go. By engaging students in a more engaging and enjoyable way, this new approach to education facilitated by the use of mobile apps is helping students to learn in a more enjoyable and effective manner.

**Common Roadblocks to Using Mobile Technology in Education Sector**

Despite its interesting features, mobile technology has been slow to catch on in the education sector due to a number of factors:

1. There appears to be an overemphasis on retroactively applying technology to problems and a reluctance to adjust where improvements are needed.

2. There is a persistent problem in the innovation process that stays in the proof-of-concept stage without progressing to scaling.

3. It seems that the inability to recognize new opportunities in mobile applications as well as the inability to update new content regularly is preventing student interest from growing.

**Overcoming Roadblocks to Mobile Technology Adoption in Education Sector**

**Building a Mobile Strategy Based on Innovation Funnel**

In order to achieve the goal of successful usage of mobile technology, it is important to first develop a mobile technology strategy that specifies clear goals and end states based on a roadmap indicating short- and medium-term objectives, along with what mobile technologies need to be employed to achieve them. Taking advantage of the heat map described above, the target end states should define technology and application objectives, and specify how a wide range of mobile learning building blocks can be applied to operational processes. It is important to examine the financial implications of the various aspects of mobile applications. In a large educational institution, a balanced portfolio of mobile projects can be developed. There are a number of parameters to consider. These include value versus feasibility, risk versus reward, scale of investment and so on. Mobile technology can be used to assist in this process, but it should be carefully chosen as a use case in order to achieve the maximum benefit. The development of an innovation funnel is necessary, which entails phases such as finding technologies and ideas to be integrated into mobile platforms, evaluating and developing them, and scaling them to integrate with offline sources in the future. Every effective innovation process relies on an open ecosystem approach, which means that external partners and users are involved in each stage of the process, such that students are an important part of the process.

**Scaling and Integrating**

Monitoring and evaluating the development of mobile technologies requires an effective scouting capability that can be implemented in real time. Mobile technology should be prioritized in line with the educational strategy and the portfolio of educational projects. It should not be merely a rigidly "top-down" exercise, but a "bottom-up" exercise that involves the end users like students in order to take advantage of new mobile opportunities that are emerging rapidly nowadays, which may not have been apparent during the process of setting the strategy.

**Assembling a Plan**

In order to move from one process stage to another, the action should be complete and explicit. The project manager should follow a transition in order to prevent a loss of product launch during subsequent activities. It is possible to accomplish focused transitions by reserving particular communication channels only to be used for transitions from one stage to the next. Interim goals are also important when making the transition from one stage to the next. The fundamental pillars for mobile technology to create value are productivity, coordination, and transformation. Without these three pillars, mobile technology is not going to do its job. In order to achieve the goal of successful mobile technology usage, and as a first step you should set up a mobile technology strategy that defines clear goals and determines what mobile technologies should be associated with achieving those goals. The roadmap should specify both short-term and medium-term goals, along with what mobile technologies will be utilized for achieving the goals. The heat map described earlier should help the target end states to identify the technology and the technology implementation objectives, as well as how a wide range of mobile learning building blocks can be utilized to support operation processes as defined by the target end states. There is a need to assess the financial implications of mobile applications in order to provide a thorough financial analysis of mobile applications.

**A Global and Local Approach**

When a mobile product is built to meet the needs of a big client with huge monetary value, when that product is introduced to a local market there tends to be a big mismatch between what engineering decided to build and what the sales team promised (or at least wanted to promise) to the client. It is very important to manage the expectations correctly in order to achieve the best results. Getting the ability to run remote product tests that validate functionality is a requirement for globally distributed local teams with different cultural and linguistic backgrounds. A failure to do so would result in the experience for local users being quite different from what they previously experienced. These challenges resulting from globalization would require functional local product test teams to meet a whole new set of requirements.

**Improve the Efficiency of Processes**

The solution to strengthening Product Launches consists of examining the current processes and to modify them so that new product launches become less painful to execute. Additionally, implementation of the proposed strategies and best practices will provide focus on the entire new product launch, from development to production. The following are five proposed practices used to strengthen product launches for Improving efficiencies from development to production.

A real time scouting capability that can be implemented in real time is essential to monitoring and evaluating the development of mobile technologies. It is important that mobile technology is prioritized according to the organizational strategy and the portfolio of projects in the educational organisation. Developing a mobile strategy should not merely be a top-down exercise, but a bottom-up exercise that involves the end users, such as students, in order to take full advantage of the new mobile opportunities that are coming up rapidly, which may not have been apparent during the process of setting up the strategy.

**Improved Project Agility**

As a rule of thumb, we should keep in mind that agility is nothing more than keeping the project open, flexible, and flexible from the start. It is for this reason that any project which deals with the launch of a mobile product needs to be flexible and that the project manager be capable of applying both a top-down and bottom-up approach to the process of planning a mobile product, the latter of which should correspond to the needs of the end users. The flexibility of the project will allow us to implement the new standards and methodologies at different points during the development of the new mobile products in a way that suits the new requirements of the project.

In order for mobile technologies to fulfill their full potential, decision makers need to focus on the efficiency and alignment of education operations. By doing so, they can begin to drive innovation, which is the key to unlocking the full potential of mobile applications. This viewpoint seeks to give the reader an understanding of the value creation process, emphasizing specifically the factors that are required in order to unlock the potential of mobile applications and systems when applied in synchrony with project agility.

**Making Decisions Based on Information**

With access to the right and high end information the project manager should make sure to refer it to subject matter experts who can help an organization correctly interpret and implement it. A strong grasp can be made as to how the operating landscape for a new product launch is standing. The information also helps in how the project manager can intelligently move forward for successfully launching a product. Most educational organizations are essentially deterministic in their approach to making decisions, relying disproportionately on structured numeric data, taken from within the educational organization, and evaluating progress against deterministic outcomes.

By combining and aggregating internal and external sources of data, it is often possible to gain new insights, intelligence, and information that provides a high return on investment. As a result, new data ecosystems may emerge, providing novel value-added services based on the data. The advent of mobile technologies may present opportunities for monetization of assets and services related to them. There may be opportunities to sell excess capacity, data, or cloud capabilities to third parties who may find them useful. As an example, within the rail industry, the commercialization of excess broadband and fiber capacity has a fairly widespread practice.

**Improving the Accuracy and Timeliness of Reporting**

There is no doubt that in today's evolving mobile landscape, performance, competitiveness, and sustainability are heavily correlated to the ability to deliver reliable results quickly and in a timely manner.

As a result of the launch of a new mobile product, one of the biggest fears that needs to be addressed is the possibility that stale reports will appear together with inaccurate statistics as to where the project stands when it comes to launch. It is necessary for the project manager to make regular visits to the development team and the routine data quality assessments must be reviewed in order to make sure that the reported data is accurate and in line with the project scope.

Production personnel have to make sure that they are well prepared to provide on-site coaching to project staff and project partners regarding data collection, accuracy and analysis in order for the action plans to be implemented in accordance with the overall vision of the project.

**Managing Constant Change and Integrating**

It is common for teams to be dispersed geographically and to work concurrently on m-commerce product launches and development processes as a result of m-commerce. Therefore, it is inevitable that software will be released at different cadences, schedules, and times. As a project manager, you may be required to get code from one team and then you may need to test the code, merge it with the code of another team, and then integrate the work of two or more teams. This process adds up to the fact that when these things are done regularly, the process of merging code has become almost a daily activity. Because of the sheer number of steps involved, this process of merging is a time-consuming one and often leads to the entire project coming to a halt as a result of the considerable amount of time and effort required. To become able to integrate all the steps involved with the developmental process in an effective way, the project manager needs a team of people who work hard and dedicate themselves to the project.

Despite the fact that project management is all about collaborative teamwork. The success of a new m-commerce product launch is directly influenced by the successful integration of individual efforts, each with a different background. A typical example of how to accomplish this would be mobile development at a small scale educational organization, which is a great example of how to assign varying levels of information access to project participants.

**Stages Leading Up to the Launch and Building Market Excitement**

The preparation process itself is one of the most exciting aspects of preparing to release a newly developed mobile app product. There are many factors that contribute to brand success and it is not just as simple as choosing the right name for the product and preparing it for success. In fact, most of these phases are just the beginning stages of the entire process. For professionals associated with the education sector, it is essential to remember that regardless of how fantastic a new app or education concept may be, it will still need to be sold. Without people knowing about the idea or concept, it will never be able to succeed. That no matter how good the idea is, it will never be able to sell. It is very important to run an enthralling campaign to raise curiosity about your app to make it a success. Even before the mobile app has been released, there may be a possibility that its reputation could be damaged if its social media presence is insufficient before the launch of the mobile app. The chances are high that your brand-new product will be put under the microscope by the internet following its launch. Consumers, industry experts, and other individuals including students regularly write about your mobile app and publish them to the public. That is why it is important for your app to have more public reviews about your products. Before a new mobile app is released on the market, a large number of ratings and reviews can be found on the Internet that have a profound impact on how it will be perceived by the public.

*Conclusion*

One of the most promising ways to deal with the present economic crisis is to develop a structured, strategic approach to mobile value creation. With the help of this study, organizations will be able to break free of the vicious circle of escalating operating costs and a lack of investment in improving mobile launches. The goal of mobile technology and apps must be to complement, rather than to replace, human efforts in order to enable students to make better and more informed decisions with greater speed and confidence, resulting in increased learning outcomes. By using mobile technology, many of the concerns in regard to its adoption are also eliminated.

In the context of an educational institution, taking a structured, strategic approach to creating value from mobile devices has great potential. It allows organizations to break out of the vicious circle of rising operating costs and a lack of investment in improving student engagement. As a starting point, mobile technology and apps should complement, rather than replace, human teaching. In this way, instructors and students can make better, more informed decisions with greater speed and confidence, resulting in greater learning. Using mobile technology within an educational setting also eliminates many of the concerns related to its adoption.

The future products which an educational sector needs to launch must develop the ability to anticipate and address customer requirements proactively, and eliminate problems preemptively wherever possible. The above listed strategies will help organizations in improving efficiencies from development to production for new m-commerce products. Reaching a level of efficiency is a constant evolutionary process that achieves total transformation founded on smarter data, smarter processes and smarter technologies. Integrating these three critical elements intelligently enables an organization to leverage innovative product launch processes, which in turn allows for the use of detailed data and technology to understand current performance and how to achieve educational goals.